\documentclass[entropy,article,submit,moreauthors,pdftex,10pt,a4paper]{mdpi} 

\usepackage{physics}
\usepackage{import}
\usepackage{textcomp}

\newcommand{\cmmnt}[1]{\ignorespaces}

\firstpage{1} 
\Title{Testing Randomness in Quantum Mechanics}

\Author{Aldo C. Mart\'inez$^{1}$, Aldo Sol\'is $^{2}$, Rafael D\'iaz Hern\'andez Rojas $^{3}$, Alfred B. U'Ren$^{2}$, Jorge G. Hirsch$^{2}$ and Isaac P\'erez Castillo$^{4,5,*}$}

\address{$^{1}$  Department of Physics, Center for Research in Photonics, University of Ottawa, 25 Templeton St, Ottawa, Ontario, K1N 6N5, Canada.\\
$^{2}$  Instituto de Ciencias Nucleares, Universidad Nacional Aut\'onoma de M\'exico, Apdo. Postal 70-543, Cd. Mx., C.P. 04510 Mexico\\
$^{3}$  Dipartimento di Fisica, Sapienza University of Rome, P.le Aldo Moro 5, I-00185 Rome, Italy\\
$^{4}$  Departamento de F\'isica Cu\'antica y F\'otonica, Instituto de F\'isica,  Universidad Nacional Aut\'onoma de M\'exico, Apdo. Postal 20-364, C.P. 04510 Cd. Mx., M\'exico\\
$^{5}$ London Mathematical Laboratory, 8 Margravine Gardens, W6 8RH London, United Kingdom
}

\corres{Correspondence: isaacpc@fisica.unam.mx}

\abstract{
Pseudo-random number generators are widely used in many branches of science, mainly in applications related to Monte Carlo methods, although they are deterministic in design and, therefore, unsuitable for  tackling fundamental problems in security and cryptography.  The natural laws of the microscopic realm provide  a fairly simple method to generate non-deterministic sequences of random numbers, based on measurements of quantum states. In practice, however,  the experimental devices on which quantum random number generators are based are often unable to pass some tests of randomness. In this review, we briefly discuss two such tests, point out the challenges that we have encountered and finally present a fairly simple method that successfully generates non-deterministic maximally random sequences.}


\begin{document}

\section{Introduction}
Monte Carlo methods are one of the essential staples of the basic sciences in the modern age. Although these gained prominence during the early 1940s, thanks to  secret research projects carried out in Los Alamos Scientific Laboratory by Ulam and von Neumann \cite{metropolis1949monte,von195113}, their origins may be traced back to the famous  Buffon's needle problem, posed by  Georges-Louis Leclerc, Comte de Buffon, in the 18th century. In the present day, Monte Carlo ``experiments'' are seen as a broad class of computational algorithms that use repeated random sampling to obtain numerical estimates of a given natural or mathematical process. 
In order to use these methods efficiently, fully random sequences of numbers are needed. Back in the 1940s this was a tall order, and various methods to generate random sequences were used (some of them literally using roulettes), until von Neumann pioneered the concept of computer-based random number generators. During the following years, these became the standard tool in Monte Carlo methods and are still generally well-suited for many applications.  However, these computer-based methods generate \emph{pseudo} random numbers, which means that the generated sequence can be determined given an algorithmic program and an initial seed, two ingredients which are  hardly random. Thus, in order to achieve a truly unpredictable source of random numbers, we must eliminate these two deterministic aspects. The former is easy to overcome using, for example, a pattern of keystrokes typed on a computer keyboard as a random seed. On the other hand, the algorithmic program could be replaced,  for instance, by a  classical chaotic system. Examples of the latter abound in the area of weather prediction and climate sciences.

In recent years, however, the community has been moving towards using the fundamental laws dictating the behaviour of the quantum realm for the generation of sequences of truly random numbers. This seems, at a first glance, to be at odds with the following rather na\"ive thought: if the natural laws of the microscopic world are considered to be a computer program under which a system evolves from an initial state (a seed), should not its corresponding generated sequence also be predictable? It turns out that Quantum Mechanics, as far as our description of this realm goes, has a special ingredient that makes the random sequence inherently unpredictable for both, the generator and the observer. Such a strange behaviour has been eloquently recast over the years in various forms, famously by the quote "spooky action at a distance" due to Einstein, or mathematically by the celebrated work of J.S. Bell  \cite{einstein1935can,bell2004speakable}. The application of quantum randomness in cryptography has given rise to the concept of device independent randomness certification which, in a nutshell, corresponds to those processes that violate Bell's inequalities \cite{pironio2010random,acin2016certified}. Yet, there seems to be some confusion in the literature regarding two different properties of a given sequence of random numbers. The first one, rather important as we have argued above, is whether the sequence is truly random, meaning that is \emph{unpredictable}. In contrast, the second one is related to the issue of assessing whether or not it is biased. It is crucial to keep in mind that these two properties are independent, as evidenced by the random number generator \textit{Quantis} \cite{quantis}, which is based on a quantum system and yet is unable to pass some tests of randomness \cite{CaludeExperimentalEvidence}.

The main goal of this review is to discuss the latter property. Although a myriad of methods exist which permit testing the quality of random number generators, for example the NIST benchmark \cite{nist2001statistical}, we  focus solely on two recently introduced approaches: the first one is based on algorithmic complexity theory evaluating incompressibility and bias at the same time, since a incompressible sequence is necessary an unbiased one \cite{calude1993borel}, while the second one relies on Bayesian model selection. We apply these methods to analyze  sequences of random bits generated in our laboratory using quantum systems. We also address the issue about the origin of the biases observed when utilizing this type of devices.

\section{Tests of randomness} 
A simple criterion for assessing the predictability of a sequence is the presence of patterns in it. For example, for the sequence $01010101...$  we can ask ourselves whether the next number is either 1 or 0. The natural answer is 0 based on the pattern observed in the previous bits. In general, we would like to find any possible regularity that helps us to {\it predict} the next bit. Within the framework of algorithmic information theory it is possible to address this problem by noting that any sequence which exhibits regularity  can be compressed using a short algorithm which can produce as output precisely such patterns. Thus, a sequence of this type could be reproduced using fewer bits than the ones contained in its original form. Therefore, whenever a sequence lacks regularity we say  it to be ``algorithmically'' random.

We now introduce a remarkable result from algorithmic information theory: the Borel-normality criterion due to Calude \cite{calude1993borel}, which allows us to asymptotically check whether a sequence is \emph{not} ``algorithmically'' random. Assuming we are given a string  $\ell=\{1001010110110\cdots\}$ of $|\ell|=n$ bits\footnote{We will only consider binary sequences, but our results are easily generalizable to other alphabets.}, the idea of the Borel-normality criterion consists primarily in dividing the original sequence $\ell$ into consecutive substrings of length $i$ and then computing the frequencies of occurrence of each of them. For brevity and later use, let us define $\Omega^{(i)} $ as the set of  $2^i$ subtrings that can be formed with $i$ characters, let   $\ell_i$ be the sequence obtained after dividing it into substrings, and $|\ell|_i \equiv [ |\ell|/i ]$.  Additionally, let $N^{j}_i(\ell)$ be the number of times  the $j$-th substring of length $i$ appears in $\ell$. For example, when considering substrings of length $i=1$ we are looking at the frequencies of the symbols $\Omega^{(1)}=\{0,1\}$ that conform the original string $\ell$, while for $i=2$, we have to consider the frequencies of four substrings, namely $\Omega^{(2)}=\{00,01,10,11\}$. According to Calude, a \emph{necessary} condition for a sequence to be maximally random is that the deviations of these frequencies with respect to the expected  values in the ideal random case should be bounded as follows \cite{calude1993borel,Calude-book}:

\begin{equation}
\left|\frac{N^{j}_i(\ell)}{|\ell|_i} -\frac{1}{2^{i}} \right|<\sqrt{\frac{\log_2 (n)}{n}}\,, \quad\quad j=0,\ldots, 2^{i}-1\,.
\label{eq:1}
\end{equation}
This condition must be satisfied for all substrings of length from $i=1$ up to $i_{\text{max}}=\log_2(\log_2(n))$. Intuitively, this criterion ``compresses'' the original sequence by reading $i$ bits at a time and tests whether the substrings appear with a frequency that differs from what would be expected in the random case, thus indicating the presence of some regularity. We emphasize that since Borel-normality is not a sufficient criterion for randomness, it can only be used to assess whether a given sequence is \emph{not} random. In other words, even if a sequence satisfies Eq.~\eqref{eq:1} for all substrings and allowed values of $i$, the Borel-normality condition cannot guarantee that it is indeed random.

Recently, a Bayesian criterion has been introduced \cite{rojas2017improving,rojas2017} by some of the authors of the present review to test, from a purely probabilistic point of view, whether a sequence is maximally random as understood within information theory \cite{cover-and-thomas}. The method works by exploiting the Borel-normality compression scheme and then, recasting the problem of finding possible biases in the sequence as an inferential one in which Bayesian model selection can be applied. Specifically, for a fixed value of $i$, we need to consider all the possible probabilistic models, henceforth denoted as $\{\mathcal{M}^{(i)}_{\alpha}\}_\alpha$, that could have generated the sequence $\ell$. Each such model determines a unique probability assignation to the elements of $\Omega^{(i)}$,  which depends on a set of prior parameters $\vb*{\theta}$. For these parameters, the Jeffreys' prior, $P_\text{Jeff}(\vb*{\theta})$, turns out to be a convenient choice of prior parameter distribution, as it entails the ``Occam Razor principle'' in which more complex models are penalized, as well as being mathematically convenient for the case at hand; some other advantages are brought forward in \cite{rojas2017improving,rojas2017}. 

Next, the question of finding all the generative models that can produce a sequence $\ell$ is ultimately solved by noticing that all the possible probabilities assignations are in a one-to-one correspondence with all possible partitions of $\Omega^{(i)}$. Since obtaining the  partitions of any set is a straightforward combinatorial task \cite{partitions}, we are able to determine all the relevant models when searching for possible biases in the generation of $\ell$. For instance, when $i=1$ there are two possible models: one in which the two elements of $\Omega^{(1)}$ are equiprobable, corresponding to the partition $\{\{0,1\}\}$ of $\Omega^{(1)}$ into one subset --\textit{i.e.} the same set-- and another model with probabilities $p_0=\theta$, $p_1=1-\theta$ corresponding to the partition $\{\{0\},\{1\}\}$ of $\Omega^{(1)}$ into two subsets. Even though it might seem that the first model is just a particular case of the second one (by letting $\theta=1/2$), we should keep in mind that the prior distributions are different in both cases, $\delta(\theta-1/2)$ and $\frac{1}{\pi \sqrt{\theta(1-\theta)}}$, respectively, thus yielding two different models. Analogously, for $i=2$ there is a single unbiased model, which corresponds to the partition of $\Omega^{(2)}$ into one subset (with probabilities $p_j=1/4$, for  $j=00,01,10,11$), and 14 additional models associated with the different ways of dividing $\Omega^{(2)}$ into subsets. The latter are related to the number of ways of distinguishing among the elements of $\Omega^{(2)}$ during the assignation of probabilities, and thus any of these models would entail some bias when generating a sequence. Note that, in general, for any value of $i$ we will face a similar situation in which a single model can produce an unbiased,  and hence maximally random, sequence by means of a uniform distribution, while the rest of them will be some type of categorical distribution.

Once all the models have been identified, the remaining part is the computation of the posterior distribution $P\left(\mathcal{M}^{(i)}_{\alpha}\Big|\ell\right)$, which from an inferential point of view is the most relevant distribution as it gives the probability that the model $\mathcal{M}_\alpha^{(i)}$ has indeed produced the given sequence $\ell$. Note that, since a generative approach was adopted, we have direct access to the distribution $P\qty(\ell\Big| \vb*{\theta}, \mathcal{M}_\alpha^{(i)})$, which can be combined with the parameters' prior $P_\text{Jeff}(\vb*{\theta})$ to obtain  the distribution $P\left(\ell\Big|\mathcal{M}^{(i)}_{\alpha}\right) = \int \dd{\vb*{\theta}} P\qty(\ell\Big|\mathcal{M}_\alpha^{(i)}, \vb*{\theta}) P_\text{Jeff}\qty(\vb*{\theta}) $. One of the most important results of \cite{rojas2017improving} is that this marginalization can be done exactly for all the models and any value of $i$. Therefore, we can obtain the posterior distribution by a simple application of Bayes' rule:
\begin{equation}
P\left(\mathcal{M}^{(i)}_{\alpha}\Big|\ell\right)=
\frac{P\left(\ell\Big|\mathcal{M}^{(i)}_{\alpha}\right)P\left(\mathcal{M}^{(i)}_{\alpha}\right)}{\sum_{\gamma}P\left(\ell\Big|\mathcal{M}^{(i)}_{\gamma}\right)P\left(\mathcal{M}^{(i)}_{\gamma}\right)}\,,
\label{eq:2}
\end{equation}
with $P\left(\mathcal{M}^{(i)}_{\gamma}\right)$ being the prior distribution in the space of models that can generate sequences of strings of length $i$. Therefore, the best model $\alpha^\star$ that describes the dataset $\ell$ is quite simply given by
\begin{equation}
\alpha^\star=\text{arg max}_{\alpha} P\left(\mathcal{M}^{(i)}_{\alpha}\Big|\ell\right)\,.
\end{equation}
If the best model $\mathcal{M}^{(i)}_{\alpha^\star}$ turns out to be the unbiased one for all possible lengths $i$ of the substrings, then we can say that the process that generated that dataset was maximally random. However, it remains to discuss how large  the length $i$ of substrings can be, for a given dataset of $n$ bits. To answer this, we first note that for any set containing $N$ elements, the possible number of partitions is given by the $N$-th Bell number $B_N$ \cite{partitions}. Thus, for a given $i$, all possible partitions of the set $\Omega^{(i)}$ will result in $B_{2^i}$ models to be tested, and, therefore, it is expected for them to be sampled at least once when observing $\ell$. This means that $B_{2^{i_{\text{max}}}}\sim n$, which, for sufficiently large $n$, yields $i_{\text{max}}=\log_2(\log_2(n))$, precisely as in the Borel-normality criterion.

Randomness characterization through Bayesian model selection has some clear and natural advantages, as already pointed out in \cite{rojas2017improving} but, unfortunately, it has an important  drawback: the number of all possible models for a given length $i$, given by $B_{2^{i}}$, grows supra-exponentially with $i$: indeed, for $i=1$ we have two possible models, for $i=2$ we have 15 possible models, for $i=3$ we have instead 4,140 possible models, while for $i=4$ we have $10,480,142,147$ models. Thus, even if we are able to acquire data for the evaluation of these many models, it becomes computationally impractical to estimate the posterior for all of them using Eq. \eqref{eq:2}. There is an elegant strategy to overcome this difficulty: one can derive bounds similar to those provided by the Borel-normality criterion, by comparing the log-likelihood ratio between the maximally random model and the maximally biased one. This yields the following bound for the frequencies of occurrences \cite{rojas2017}:
\begin{equation}
\sqrt{\sum_{j\leq j'=1}^{2^i -1}\left(\frac{N^{j}_i(\ell)}{|\ell|_i}-\frac{1}{2^i}\right)\left(\frac{N^{j'}_i(\ell)}{|\ell|_i}-\frac{1}{2^i}\right)}<
\sqrt{\frac{i^2}{n^2\psi_1\left(\frac{1}{2}+\frac{n}{i2^i}\right)} 
\ln\left(\frac{2^{-n}\Gamma^{2^i}\left(\frac{1}{2}\right)\Gamma\left(\frac{1}{2^{1-i}}+\frac{n}{i}\right)}{\Gamma\left(\frac{1}{2^{1-i}}\right)\Gamma^{2^i}\left(\frac{1}{2}+\frac{n}{i 2^i}\right)}\right)}\,,
\label{eq:3}
\end{equation}
where $\psi_1$ is the polygamma function of order 1. Note that, unlike Calude's bound given by Eq. \eqref{eq:1}, this new Borel-type bound couples all frequencies, and moreover, results into highly restrictive bounds.

\section{Ideal random number generation}

While intuition dictates that quantum random number generators (QRNG) should be superior to their classical counterparts, such a comparison was  carried out in \cite{CaludeExperimentalEvidence} and very recently in \cite{abbott2018experimentally}, with rather disappointing results. For the classical case, the authors used three Pseudo-Random Number Generators (PRNG): the generators included in the software packages Mathematica and Maple, and the digits of $\pi$ expressed in base 2. For QRNG they used two devices: i) Quantis,  developed by IDQ \cite{quantis}; a quantum random number generator interfaced with a common computer, and ii)  an experiment from a quantum optics group in Vienna. The latter experiment consists of a very weak light source, attenuated to the single photon level, a beam splitter, and two single photon detectors. Leaving aside the question of which QRNG performs better, the real surprise was that the PRNGs come out in this test with a superior performance, by far,  as compared to their quantum counterparts. This result appears to be  at odds with the natural  randomness associated with quantum phenomena.  Why is it that the inherent quantum randomness does not translate into better performance with respect to classical systems? Does randomness, as discussed in this paper,  have no impact on the performance of the generators? Is this a fundamental or a technical problem?\\

\begin{figure}
  \centering
  \includegraphics[width=6cm]{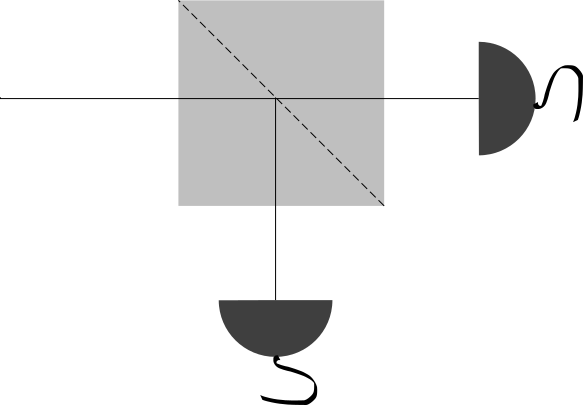}
  \caption{Ideal experimental setup for a na\"ive QRNG based on an individual photon source and a beam splitter (BS). Neglecting the possible losses, a photon will activate only one of the two detectors, therefore producing a random bit per photon.}
  \label{fig:beam-splitter}
\end{figure}

In order to explore this apparent paradox  we will discuss the different technical and design difficulties associated with quantum random number generation using light.   These days it is straightforward to detect single photons using avalanche photo-diodes (APD), devices capable of detecting up to a few million single photons per second with  $>60\%$ detection efficiencies employing relatively simple electronics.  With this simple design in mind, we only need a single photon source, a  beam splitter (BS), and a couple of single-photon detection devices in order to set up a QRNG device.  This minimalistic design  is sketched in Figure \ref{fig:beam-splitter}.

\section{Experimental challenges in random number generation}
Suppose now that we have a single-photon source and  we want to generate a sequence of bits to be tested against the bounds given by Eq. \eqref{eq:1}. Let us start by first focusing  on the so-called Borel level, the word length $i$, which can take a maximum value of $i_\text{max}=\log_2(\log_2(n))$. In table \ref{table:nlist} we report how $i_\text{max}$ grows with $n$, up to a value of $i_\text{max}=6$. In order to achieve a Borel level  $i_\text{max}=6$ we will require a dataset of length $10^{18}$ events. Assuming that our single-photon generator and  detectors can cope with around a million of events per second, we would then require of the order of $600$ years to generate a sequence of that length!   It turns out that $i_\text{max}=5$ is a more realistic value, since it leads to a required  dataset $\ell$ of size $4.3\times10^9$ events, which can be realistically produced in  a couple of hours.
\begin{table}
\centering
\begin{tabular}{||c|c||}
	\hline
  Maximum Borel level & data length \\
	$i_{\text{max}}=\log_2(\log_2(n))$ & $n$ \\
    \hline
    1 & 4 \\
    2 & 16 \\
    3 & 256 \\
    4 & 65,536 \\
    5 & 4,294,967,296 \\
    6 & 18,446,744,073,709,551,616 \\
    \hline
\end{tabular}
\caption{Necessary data lengths for maximum Borel level $i_{\text{max}}=\log_2(\log_2(n))$. The double exponential relation grows so quickly that it is not possible get to level 6. }
\label{table:nlist}
\end{table}

The bound on the right-hand-side of Eq.~\eqref{eq:1} implies that the frequencies of occurrences $\frac{N^{j}_i(\ell)}{|\ell|_i}$ for  substrings of length $i\leq i_{\max}$ cannot deviate from the ideal random one, $1/{2^{i}}$, by more than  $8.6\times 10^{-5}$, which constitutes an extremely tight tolerance.  Hence, in practice, any part of the na\"ive experimental setup that gives rise to some bias, will unfortunately make the dataset $\ell$ unable to pass the Calude criterion.  The first component that we must be wary of is the BS. A regular BS usually has an  error figure in the region of 1\%, which is very high with respect to the stringent tolerance $\ell$ would need to pass for Borel normality. Is it plausible to correct this using a Polarizing Beam Splitter (PBS), instead of the BS, with an active control through feedback of the state of polarization so as to compensate for any bias in the PBS?   In what follows we investigate this question through a simple experiment.  The state of polarization of a single photon entering the PBS can be written as

\begin{align}
	\ket{\psi} = a\ket{V}+e^{i\phi}b\ket{H},
  \label{eq:state}
\end{align}

\noindent where $\ket{V}$ and $\ket{H}$ refer to the vertical and horizontal polarization components, respectively.    We can approach the state in Eq. \eqref{eq:state} by transmitting the laser beam trough  a half wave plate (HWP)  so as to achieve arbitrary rotation of the linear polarization.   Assuming a perfect, unbiased BS, we would need an incoming polarization state with $a=b=1/\sqrt{2}$ so that the resulting sequence of bits is unbiased. If, on the other hand, the PBS exhibits biases (e.g. due  to manufacturing error), we can adjust the orientation of the above-mentioned half wave plate so as to adjust precisely the value of our coefficients $a$ and $b$ to compensate for the PBS bias.

\begin{figure}
  \centering 
  \includegraphics[width=0.8\textwidth]{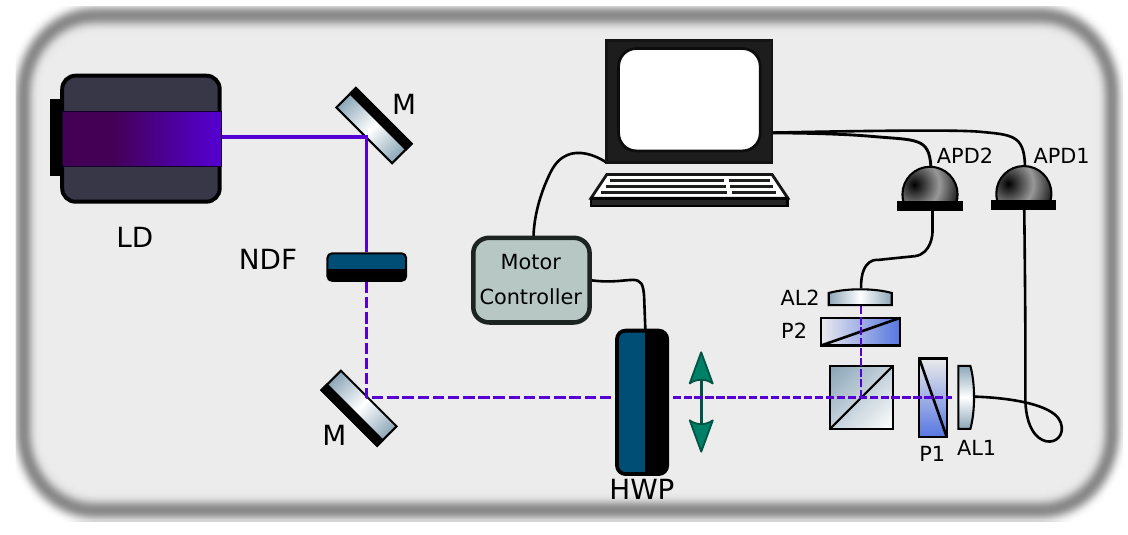}
  \caption{Experimental setup. The relative angle of the half wave plate is controlled in order to reduce bias.}
  \label{fig:experimental}
\end{figure}

Our experimental setup, shown in Figure \ref{fig:experimental}, can be regarded as the minimal realistic device for the implementation of a QRNG. The main questions which we wish to address are: i) how good are the sequences of bits generated by such a device? And ii) do they pass the Borel-normality criterion?   

The input state is prepared using the beam from a laser diode (LD). The beam is transmitted through a set of neutral density filters (NDF) with a combined optical density $7.3$ for  attenuation to a level compatible with the maximum recommended count rate of our single photon detectors.   The beam is then transmitted through a  half wave plate (HWP) mounted on a motorized rotation stage so as to control its orientation angle relative to the PBS axes.    The PBS splits the beam into two spatial modes according to the H and V polarizations, each of which is coupled with the help of an aspheric lens (AL1 and AL2) into a multimode fiber leading to an avalanche photodiode (APD1 and APD2).   We include a polariser (P1 and P2), with an extinction ratio, defined as the ratio of the maximum to the minimum transmission of a linearly polarized input,  of $100,000:1$  prior to each of the aspheric lenses (AL1 and AL2) for a  reduction of the non-polarized intensity reaching the detectors.

\begin{figure}
	\centering
  \includegraphics[width=0.9\textwidth]{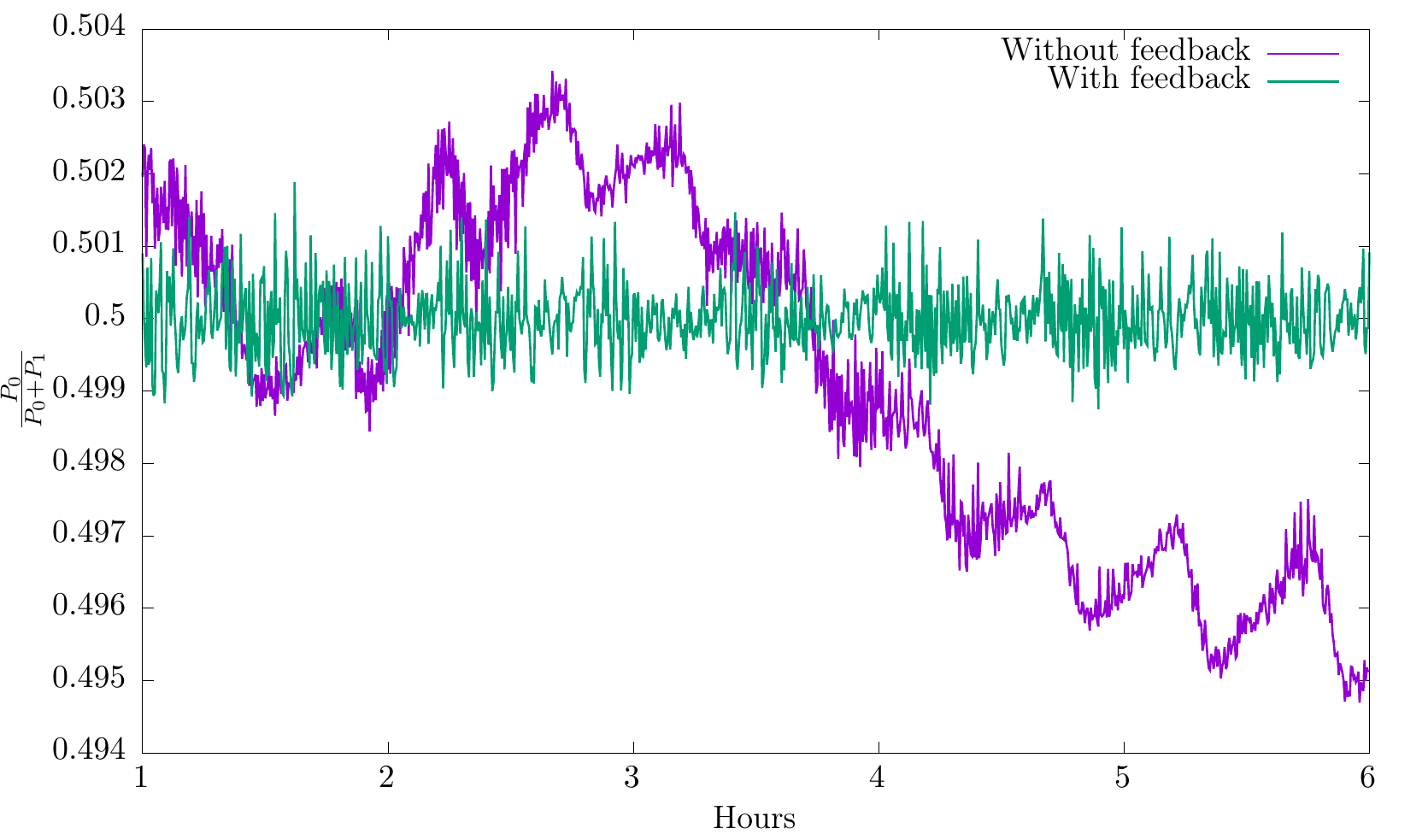}
    \caption{Evolution in time of the normalized power. The device starts in a perfect balanced state but quickly deviates from this condition. By using a feedback mechanism we can obtain stability of the normalized power within an error of  $0.004$. }
    \label{fig:timeEvolution}
\end{figure}

Suppose now  that we  prepare our system so that the average relative power $P_i/(P_0+P_1)$ of each APD detector $i=0,1$ starts ideally at $1/2$. In Figure \ref{fig:timeEvolution} we show how this  average relative power evolves with time (see curve labelled 'without feedback').    Note that, even though the system starts in a perfectly balanced state, it rapidly deviates from this condition. This is probably due to thermal drift, which can be also compensated by rotating the HWP.  After some study of the response function of our experimental setup, a correction every minute with a proportional controller was sufficient to correct for all these effects and get a steady response (see curve labelled 'with feedback').

\section{First battery of results}
\begin{figure}
  \centering
   \includegraphics[width=0.99\textwidth]{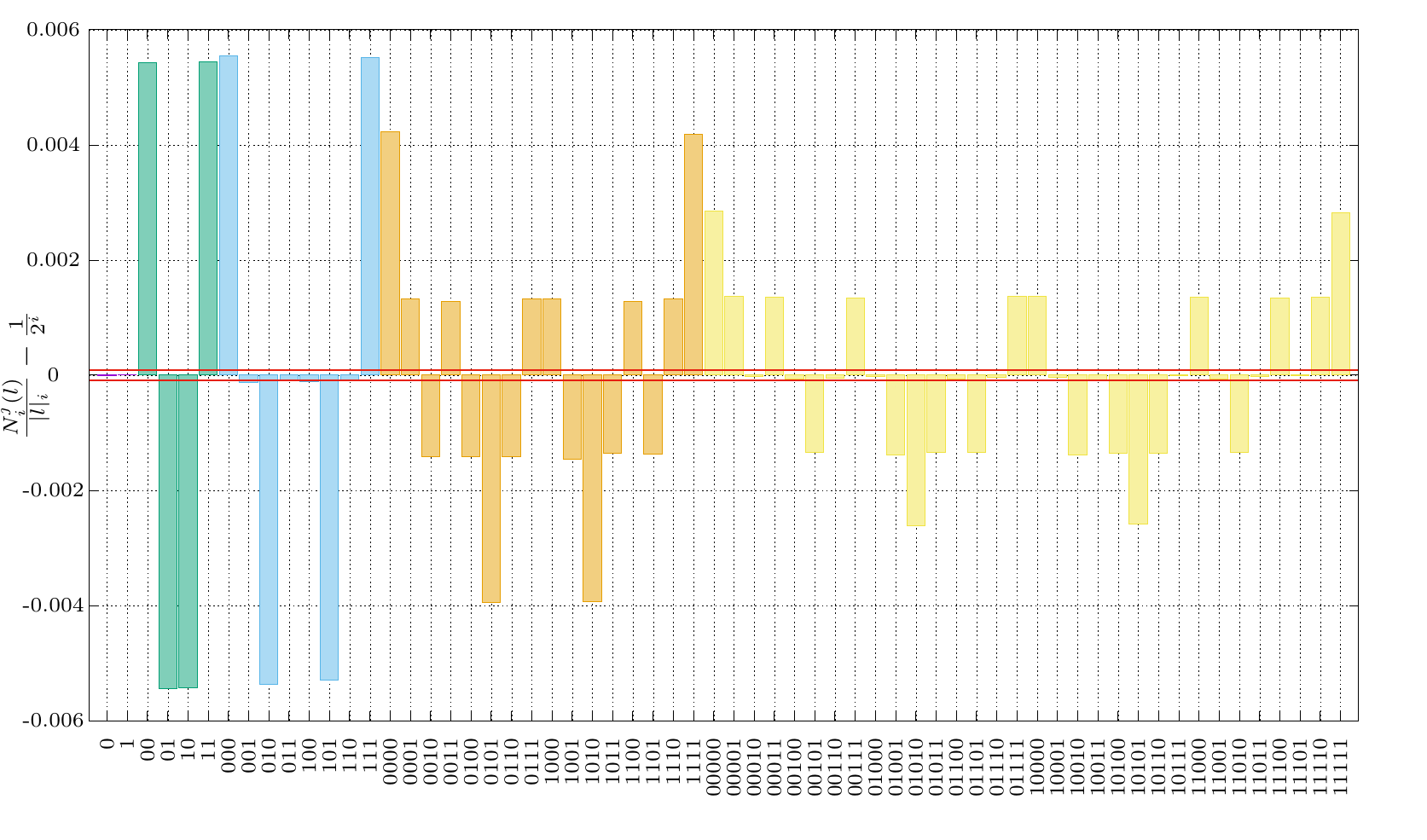}
  \caption{Results from Borel analysis. The first two boxes correspond to the deviations from the mean at the first Borel level; these exhibit the same height but opposite signs. The next four bars (green) represent the deviations for level two, i.e. '00','01','10','11'. The blue and yellow boxes represent the deviations for level three and four, respectively. The red lines correspond to  Borel's bound, which turns out to be much smaller than the deviations.  Only the first level passes the test.}
  \label{fig:res-borel-2}
\end{figure}

We have used the experimental setup described in the previous section to generate a sequence of $4,294,967,296$ bits, allowing us to  test the Borel-Normality criterion up to level $i_{\text{max}}=5$. The results of this analysis are depicted in Figure \ref{fig:res-borel-2}. In the plot, bars represent the deviations from the ideal value for all the strings at each Borel level. For instance, in the first part of the analysis (purple bars) there are only two bars corresponding to the frequency of occurrences of substrings $'1'$ and $'0'$.  As  our initial  setup is very fine-tuned and stable, the bars have practically zero height, with value $5\times10^{-6}$.  The green bars represent the second part of the analysis, or Borel level two, corresponding to frequencies of occurrences of symbols $\{00,01,10,11\}$, and so on. In the same figure, the horizontal lines represent  the bound given by the right-hand-side of  Eq. \eqref{eq:1}. Our first battery of results are a clear disappointment: only the first set for substrings of length one clearly passes the test, while for  higher lengths, our QRNG fails miserably to pass Calude's criterion. 

Furthermore, a closer look at the green bars shows that events 00 and 11 appear more frequently than expected, by about $0.005\%$, (while events 01 and 10 appear less frequently than expected by the same margin). This effect reveals a correlation between `equal events', that is, the same digit appearing twice.  In terms of our experiment, this means that it is more probable to observe an event in a detector once a previous event has already been recorded.  Other parts of our test validate this: for Borel level three the yellow bars indicate that events with alternate zeroes and ones (010 and 101) appear less frequently than expected, also by about $0.005\%$. At Borel levels four and five, the larger deviations appear for events 0101 and 1010, clearly in accordance with the previous results. This indicates that certain parts of our experimental setup are introducing unwanted correlations between bits which results in the magnitude of some of the deviations  to be 50 times larger than expected. Our experimental effort clearly does not suffice for our sequences to pass the Borel-normality criterion. How is this possible?

\section{APD effects on introducing correlations}

The two main effects in the behavior of our APDs which can introduce undesired correlations in the resulting sequences of bits are called  \textit{after-pulsing} and \textit{dead time} \cite{afterpulsing,dark-counts}. The first effect, roughly speaking, corresponds to a false detection event due to the residual effects of an avalanche  triggered by a previous 
event, while the dead time is the time period after each event during which the system is not able to record a subsequent incoming optical signal. 

The mechanism by which dead time introduces correlations in our data, particularly in experimental arrangements with two or more APDs as in our case, is as follows: suppose that we have an event in one of our detectors. During its dead time it will have zero probability of recording another incoming event in that detector, while the other detector still exhibits a non-zero probability of recording an event.  On the other hand,  the way after-pulsing introduces correlations in our data is by increasing the probability of observing consecutively the same event, resulting in the observation of an excess of the events $\{00\}$ and $\{11\}$ in Fig. \ref{fig:res-borel-2}.\\
These two effects can somewhat be corrected either by re-designing our experimental setup and/or modifying the software. Instead of following this route to generate maximally random sequences of bits,  let us pursue a rather simple solution as discussed below.

\section{Random number generation using time measurements}
We now follow a method introduced in \cite{isida1956random}. Suppose that $\rho(x)$ is the probability density function of a continuous random random variable $X$ on an interval $x\in(a,b)$. Let us further assume that its real value $x$ is represented up to a given precision so that we assign a parity to $x$ according to the parity of its least significant digit. Next, we divide the interval $(a,b)$ into an even number $2L$ of bins and introduce
\begin{equation}
x_{i}=a+\frac{i(b-a)}{2L},\quad\quad i=0,\ldots,2L\,.
\end{equation}
Suppose that $2L$ and the precision has been chosen so that $x_{i}$ is even for $i$ even and odd for $i$ odd. It follows that
\begin{equation}
\begin{split}
1&=\int_a^b dx \rho(x)=\sum_{i=0}^{2L-1}\int_{x_i}^{x_{i+1}} dx\rho(x)\equiv \mathcal{N}_{\text{even}}+\mathcal{N}_{\text{odd}}\,,
\end{split}
\end{equation}
with
\begin{equation}
\mathcal{N}_{\text{even}}\equiv \sum_{i=0}^{L-1}\int_{x_{2i}}^{x_{2i+1}} dx\rho(x)\,,\quad\quad \mathcal{N}_{\text{odd}}\equiv \sum_{i=0}^{L-1}\int_{x_{2i+1}}^{x_{2i+2}} dx\rho(x)\,.
\end{equation}
Approximating the integrals by the left sum rule we can write that
\begin{equation}
\mathcal{N}_{\text{even}}\sim \frac{b-a}{2L} \sum_{i=0}^{L-1}\rho(x_{2i})\,,\quad\quad \mathcal{N}_{\text{odd}}\sim \frac{b-a}{2L} \sum_{i=0}^{L-1}\rho(x_{2i+1})\,,
\end{equation}
which implies that, roughly, the probability that the least significant digit is odd can be expressed as
\begin{equation}
\mathcal{N}_{\text{odd}}\sim\frac{1}{2}+\frac{1}{2}\sum_{i=0}^{L-1}\rho'(x_{2i})\left(\frac{b-a}{2L}\right)^2\,,
 \end{equation}
 where the bias term can be fine-tuned by increasing either the number of bins or through a smooth density $\rho(x)$, or both.\\
 This method can be very easily implemented in the lab as follows. Suppose that  the random variable $X$ is the time difference between two consecutive photon arrivals to the detector.  In our case, these times are of the order of $500$ ns to $10$ \textmu s. A typical sequence of these time differences look like:
 \begin{align*}
592\ 342 ps \\
595\ 634 ps \\
593\ 645 ps \\
592\ 342 ps \\
595\ 634 ps \\
    \vdots
\end{align*}

\begin{figure}[h!]
  \centering
   \includegraphics[width=0.99\textwidth]{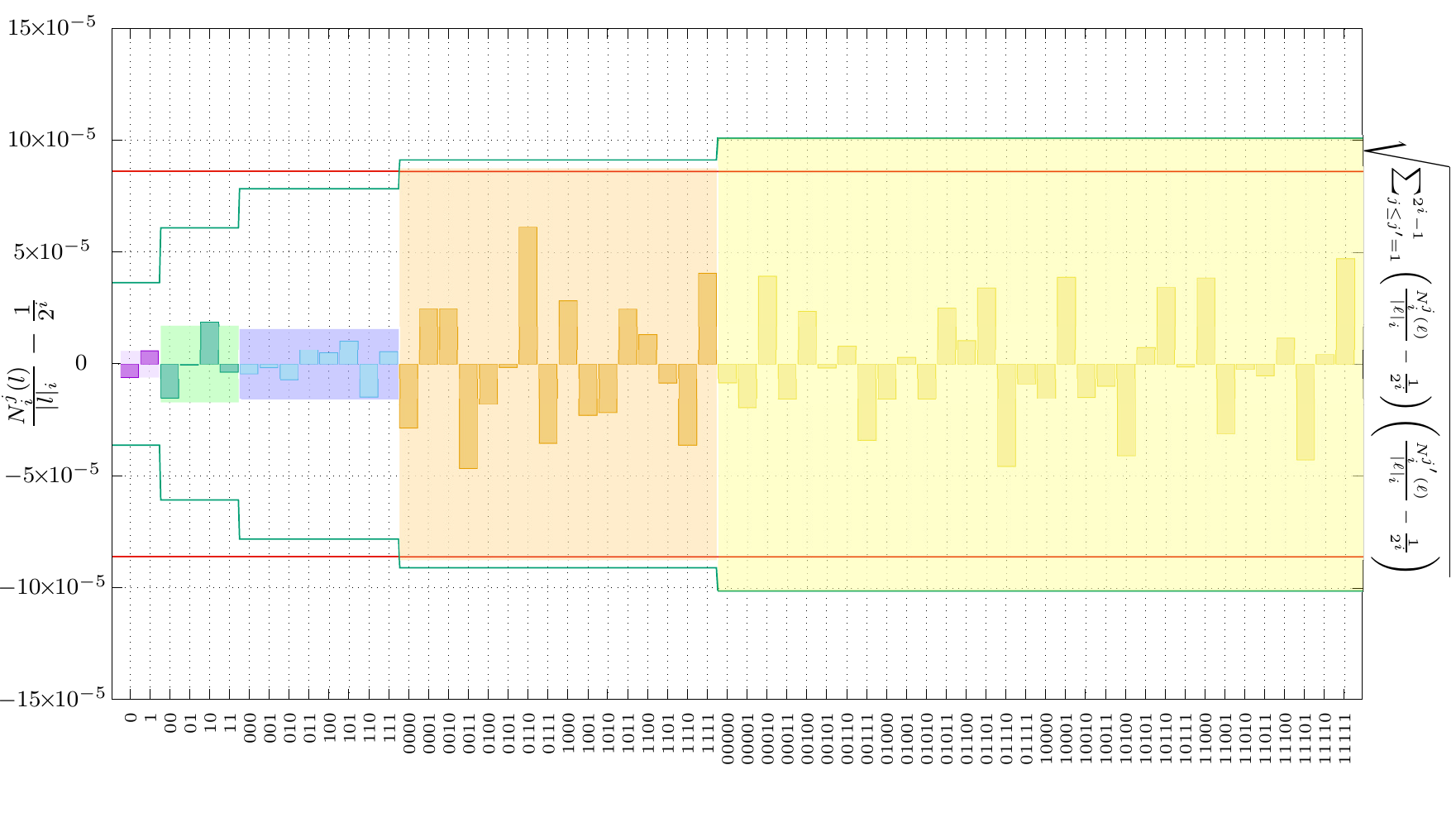}
  \caption{Results for generation using the least significant bits of time tags. In this case the deviations are very small, so this generation scheme is excellent. The solid red lines represent the maximum deviations allowed by the Calude test, while the solid green lines corresponds to the bounds given the Bayesian approach. }
  \label{fig:pasan}
\end{figure}

We can then look at the parity of the least significant digit and assign, for instance, a 0-bit to even parity  and 1-bit to odd parity, thus generating a dataset $\ell$ of $n$ bits. In Fig. \ref{fig:pasan} we show the results of testing such a sequence using the Borel-normality bounds and Bayesian bounds, given by Eqs. \eqref{eq:1} and \eqref{eq:3}, respectively, up to Borel level $i_{\text{max}}=5$.  The colored bins correspond to the deviations from the ideal value of relative frequencies  for all the possible subsequences. These  are ordered using its length $i=1$ (purple bins), $i=2$ (green bins), $i=3$ (blue bins), $i=4$ (orange bins), and $i_{\text{max}}=5$ (yellow bins). The solid red line corresponds to the Borel bound, $8.6\times 10^{-5}$. In the same graph the green line is the Bayesian bound given the right-hand-side of Eq. \eqref{eq:3} that depends  on $i$ and therefore it is not a constant, as is the case for the Borel bound. Finally, the height of the various background colored boxes correspond to the values given by the left-hand-side expression of the Bayesian bound.

As we can see, this extremely simple QRNG passes the Borel-normality criterion up to $i=5$, and nearly passes the Bayesian criterion (passes it for $i\le 4$ and slightly exceeds the bound for $i=5$).   Notice that, while the previous experimental setup required an accurate balance between zeroes and ones, in the present case we have already very small deviations at Borel level $i=1$, less than $10^{-5}$, showing the convenience of this method. For  $i=2$ (green boxes) the deviations are much larger, almost three times the value for $i=1$, but nevertheless they pass the test again by a considerable margin. These results show the lack of correlations between consecutive events, which is the main drawback of the previous approach.  It is important to note that in the results at Borel level $i=4$  there is  a substantial increase in the deviations compared with $i=1,2,3$. While this increase may indicate the presence of some as yet unidentified correlations, these  are of an insufficient magnitude to reach the bounds. On the other hand this experimental setup fails to pass some of the requirements of the Bayesian scheme. The deviations derived from the Bayesian criterion are shown in Table \ref{table:Bayesian}, and also  in Fig. \ref{fig:pasan}. As we can see, all Borel levels pass the Bayesian test, except for the last one, albeit by a small margin.

\begin{table}
\centering
\begin{tabular}{||c|c|c|c||}
	\hline
  Borel level & LHS of Eq. \ref{eq:3}&RHS of Eq. \ref{eq:3}& LHS<RHS \\
    \hline
    1 &5.719$\times10^{-6}$& 3.62956$\times10^{-5}$&Yes  \\
    2 &1.7129$\times10^{-5}$& 6.08097$\times10^{-5}$& Yes\\
    3 &1.54974$\times10^{-5}$& 7.82572$\times10^{-5}$&Yes \\
    4 &8.74186$\times10^{-5}$& 9.11726$\times10^{-5}$ & Yes\\
    5 &1.01138$\times10^{-4}$& 1.01069$\times10^{-4}$ &No \\    \hline
\end{tabular}
\caption{Comparison between the left-hand-side and  the right-hand-side of the Borel-type bounds given by Eq. \eqref{eq:3}, applied to the sequence of bits generated in our lab.  As we can observe, the Bayesian bounds are satisfied for the first four Borel levels, but not the last one by a slight margin.}
\label{table:Bayesian}
\end{table}

We can also look at the value of the posterior distribution, given by Eq. \eqref{eq:2} for the maximally random model. The value of the posterior for the four word lengths is reported in Table \ref{table3}. For the first three Borel levels, the posterior distribution of the maximally random model $\alpha_{\text{sym}}$  is very close to one, indicating that, given the dataset, this is the most likely model to have generated such data. For Borel level $i=4$, we are only able to analyse those models which are in the vicinity, in parameter space, to the maximally random model. These models correspond to partitioning $\Omega^{(5)}$  into two subsets, resulting into $32,767$ models, giving a total of $32,768$, including the maximally random one. In this case, it turns out that the most likely model is not the maximally random one. Actually, using the value of the posterior probability, this model is ranked in the position 9,240 out of all the explored models, and therefore the sequence of bits fails to pass the Bayesian criterion already at Borel level $i=4$.  Note that $i=5$ was not included in Table \ref{table3}, because we lack the computational power to address this Borel level. 

\begin{table}
\centering
\begin{tabular}{||c|c|c||}
	\hline
  Borel level $i$&Number of models  analysed  &$P\left(\mathcal{M}^{(i)}_{\text{sym}}\Big|\ell\right)$ \\
    \hline
    1 &$B_{2^i}=2$ &0.999984 \\
    2 &$B_{2^i}=15$&0.999634\\
    3 &$B_{2^i}=4,140$ &0.995476\\
    4 &32,768 models considered out of $B_{2^i}=10,480,142,147$&9.2179$\times10^{-42}$\\   \hline
\end{tabular}
\caption{Value of the posterior distribution $P\left(\mathcal{M}^{(i)}_{\text{sym}}\Big|\ell\right)$ given by Eq. \eqref{eq:2}  for the maximally random model. Note that the prior distribution for each Borel model is a flat distribution along all the models tested. This means, for instance, that at level $i=3$, while we do not have any observational bias to choose among any particular model, that is $P(\mathcal{M}_\alpha^{(i)})=\frac{1}{4140}$, after observing the data, the maximally random model is the most plausible.}
\label{table3}
\end{table}

\section{Conclusions}

A vast literature exists which claims that QRNGs are superior when compared to their classical counterparts, based on purely theoretical arguments. Indeed,  randomness in Quantum Mechanics is usually justified by the unpredictability of  individual measurement outcomes given some initial conditions. More concretely,  quantum unpredictability is based on no-go theorems, such as Bell's theorem, that simply tells us that given some initial conditions it is impossible to predict the outcome of a single measurement. However, in the present review we have shown  that QRNGs actually perform rather poorly in tests of randomness as compared to classical PRNGs.  The reason is fairly simple: unpredictability has nothing to do with bias, and while experimental devices based on Quantum Mechanics may produce a truly unpredictable random signal, they also tend, more often than not, to introduce correlations. In particular, for QNRGs based on optical devices, we have been able to account for two,   perhaps amongst the many, effects that introduce bias in our data.    While in our own experimental work involving a QNRG we have failed to obtain sequences which obey the Borel and Bayesian criteria, we were able to show that extracting sequences from the least significant digits of times of arrival represents a promising strategy.



\acknowledgments{
Financial support of  UNAM-DGAPA-PAPIIT IA103417 and IN109417 is acknowledged.
}

\authorcontributions{J.H and A.R. conceived and designed the experiments; A.M. and A.S. performed the experiments; I.P.C. and R.D.H.R. developed the method based on Bayesian Inference and analysed the data. All authors contributed to writing the paper.}

\conflictsofinterest{The authors declare no conflict of interest.} 

\appendixtitles{no} 
\appendixsections{multiple} 

\bibliography{bib}

\end{document}